\documentclass[%
 reprint,
 amsmath,amssymb,
 aps,
 prl,
 twocolumn,
 superscriptaddress,
]{revtex4-2}

\usepackage{graphicx}
\usepackage{dcolumn}
\usepackage{bm}

\usepackage{xcolor}
\usepackage{float}

\begin{document}


\title{Dancing dervishes, spinning skirts and whirling waves}

 \author{Paul R. Kaneelil}
    \affiliation{John A. Paulson School of Engineering and Applied Sciences, Harvard University, Cambridge, MA 02138, USA.}
 \author{Sifan Yin}
    \affiliation{Max Planck Institute for the Physics of Complex Systems, 01187 Dresden, Germany.}
 \author{L. Mahadevan}\email{lmahadev@g.harvard.edu}
    \affiliation{John A. Paulson School of Engineering and Applied Sciences, Harvard University, Cambridge, MA 02138, USA.}
    \affiliation{Department of Organismic and Evolutionary Biology, Harvard University, Cambridge, MA 02138, USA.}
    \affiliation{Department of Physics, Harvard University, Cambridge, MA 02138, USA.}

\date{\today}

\begin{abstract}
Inspired by the patterns seen in the tunic of a whirling dervish or the long skirt of a Kathak dancer, we study the emergence of dynamical patterns in a \emph{soft} heavy spinning elastic sheet. Using experiments, we show that the initially draped sheet can pass through three archetypal states as it rotates: a gravity-dominated draped rotation state, a flat state, and finally a whirling state marked by retrograde rotation of wavy patterns relative to the rotating frame. A minimal theory shows how gravity-induced curvature renormalizes the elastic stiffness and gives rise to a whirling resonance frequency that is a fractional-power geometric mean of elastic and gravitational frequency scales, and is independent of the sheet thickness in the gravity-dominated limit. Our experiments and theory yield a phase diagram for the dynamical states of spinning sheets, illustrating how externally induced geometry can qualitatively transform the stability of elastic continua.  

\end{abstract}

\maketitle

\begin{figure}[b]
\begin{center}
\includegraphics[width=0.39\textwidth]{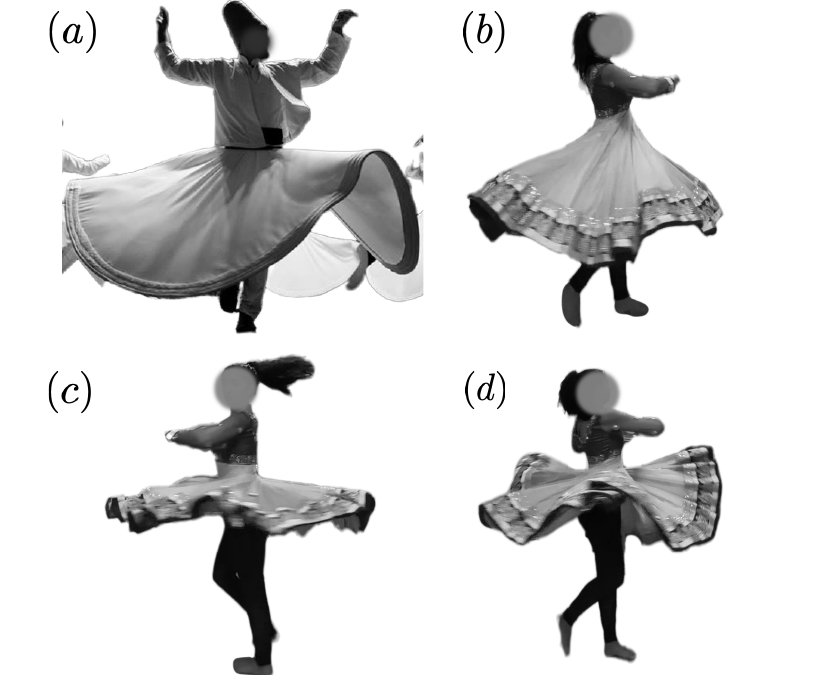}
\caption{\label{figshapes} Patterns observed on the tunic of a (a) whirling dervish~\cite{istanbuldervishceremony_sirkeci} and on the skirt of a Kathak dancer [Credit: Aamy Kuldip] spinning at different tempos: (b) low tempo ($\Omega \approx 0.7$ Hz), (c) to medium, (d) and to high ($\Omega \approx 1.33$ Hz).}
\end{center}
\end{figure}

The aesthetic shapes of static draped sheets are determined by the competition between gravity and elasticity and underlie the fluted patterns seen in table cloths,  hanging curtains, skirts, and saris \cite{cerda2004elements,boedec2021disk}. The dynamics of drapes associated with the tunic of a whirling dervish, the clothes on a model on a fashion walk, or the skirt of a dancer spinning to music are arguably even more beautiful, as shown for example in Fig. \ref{figshapes} (see also Refs. \cite{davidsbeenhere_sufi_dervishes,chakraborty2022taaldhamaar}). 
From a physical perspective, the subtle interplay of gravity, inertia, elasticity, and aerodynamic forces inherent in these systems raises natural questions of the stability of these morphologies and the regimes of their (co)existence. Prior work on rotating sheets has focused mainly on the transition from a buckled to an axisymmetric state driven by centrifugal stresses  \cite{simmonds1962finite,okuizumi2007equilibrium,chen2011warping,okuizumi2014forced,delapierre2018wrinkling,coman2023wrinkling}, traveling waves in conical sheets \cite{guven2013whirling}, and on instabilities induced by fluid stresses~\cite{yin2026elastohydrodynamic}. In this Letter, we study the dynamical steady states and the transitions seen in spinning skirts and tunics using a minimal experimental mimic - a heavy soft spinning sheet clamped at its center. 

Rotation generates centrifugal tension that generally suppresses transverse deformations in elastic sheets. Can increasing rotation nevertheless lead to the reemergence of nonaxisymmetric shapes, and can these shapes rotate independently of the material? Furthermore, how does the curvature of an initially draped sheet modify the classical dynamics of a rotating plate? Here we show that the interplay of gravity, geometry and inertia leads to a reentrant breaking of rotational symmetry and a gravity-modified criterion for whirling that is qualitatively different from its classical counterpart. 
More specifically, we find that the emergent steady-state shapes can be classified into three archetypal states: a gravity-dominated draped rotation state, a flat state, and a whirling state in which an up-down symmetric shape of the sheet undergoes retrograde rotation, where the shape rotates slower than the imposed spin. All three states are realized by a Kathak dancer spinning at different tempos, as shown in Fig. \ref{figshapes}(b)-(d), although we note that the dancer is not at a perfect steady-state.

\begin{figure*}[t]
\begin{center}
\includegraphics[width=0.99\textwidth]{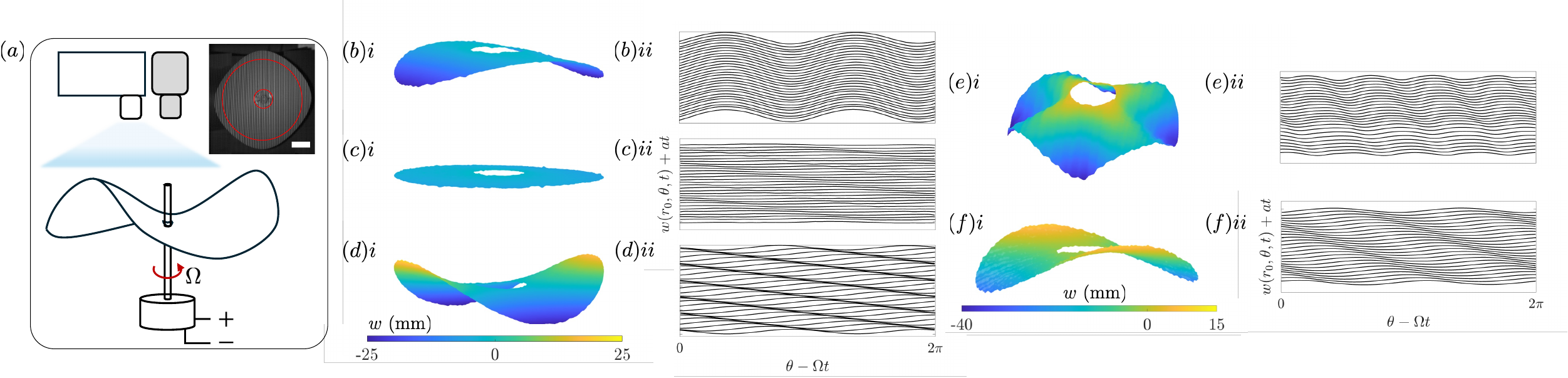}
\caption{\label{figsetup} Three dimensional reconstruction of the shape of the elastic sheet. (a) Schematic of the Fourier Transform Profilometry technique, where a fringe pattern is projected on to the elastic sheet and is imaged by a high speed camera placed beside the projector. Inset shows a representative top view image from the high speed camera showing the fringe pattern on a deformed sheet. Scale bar represents 4 cm. (b-d) Representative reconstructed shapes ($i$) and associated kymographs ($ii$) showing Sequence 1: (b) draped rotation, (c) flat, and (d) whirling. Kymographs show the spatiotemporal evolution of the profile of the sheet at a fixed radius $r_0$ offset by an arbitrary constant $a$ in order to clearly display the wave propagation. (e-f) Representative reconstructed shapes ($i$) and associated kymographs ($ii$) showing Sequence 2: (e) draped rotation and (f) whirling.}
\end{center}
\end{figure*}
To reproduce these patterns in the laboratory, we carried out experiments using laser cut  sheets of high density polyethylene (HDPE; McMaster-Carr) with a thickness $h=0.42$ mm, density $\rho=968.8$ kg/m$^3$, Young's modulus of $E=1.3$ GPa (measured with a dual-column Instron), and Poisson's ratio of $\nu=0.46$ \cite{hdpeproperties}, natural rubber (McMaster-Carr) with $h=1.57$ mm, $\rho=920$ kg/m$^3$, $E=1.9$ MPa, and $\nu = 0.5$, or custom-made sheets of vinyl polysiloxane (VPS; Zhermack 8 shore A) cast in  3D printed molds with $h=[0.5,~0.65,~1.3]$ mm, density $\rho=1070$ kg/m$^3$, $E=0.2$ MPa, and $\nu = 0.5$, with the bending stiffness $D=Eh^3/12(1-\nu^2)$. The radii of the sheets were in the range $R \in [5,~12]$ cm. All sheets were clamped at the center with an inner radius $R_i=1.6$ cm, and connected to the shaft of a 9V DC motor. The input voltage was controlled to achieve different angular speeds. To quantify these shapes, we used Fourier Transform Profilometry (FTP) \cite{takeda1983fourier,cobelli2009global,hartong2014experimental}  with a projector (Yaber Y30) and camera (Phantom v7.3)  as shown schematically in Fig. \ref{figsetup}(a). 
Post-processing the images gives the 3D reconstructed shape, shown in Fig. \ref{figsetup}(b-f). The deformed shape of the sheet is described by $w(r,\theta,t) = w_0(r,t) ~\textrm{cos}[m(\theta-\Omega_s t)]$, where $w_0(r,t)$ is the amplitude and $\Omega_s$ is the angular speed of the pattern, with $\Omega_s=\Omega$ when the pattern co-rotates with the shaft. We note that only an annular region (see   Fig. \ref{figsetup}(a) inset) of the actual sheet is reconstructed,  since including the edges lead to imaging discontinuities.

As $\Omega$ is increased, the shape of a given sheet evolves in one of two sequences: (1) draped rotation, to flat, to whirling or (2) draped rotation directly to whirling. Figure \ref{figsetup}(b)$i$ shows a representative image from an experiment showing the draped rotation state for sequence 1 and Fig. \ref{figsetup}(b)$ii$ shows the associated kymograph, where the x-axis is the azimuthal coordinate in the rotating frame, and the y-axis is the deflection at a fixed radial location offset over time. Note that the shapes align vertically in the kymograph, indicating a rigid body rotation with the rotating shaft. As $\Omega$ increased, the sheet flattened more and more until it reached a fully flat state. Figure \ref{figsetup}(c)$i$ shows an experimental image of the flat state, along with a corresponding kymograph in Fig. \ref{figsetup}(c)$ii$. Increasing $\Omega$ further, the sheet transitioned to whirling and exhibited a large amplitude deflection, as shown in Fig. \ref{figsetup}(d)$i$. The corresponding kymograph in Fig. \ref{figsetup}(d)$ii$ shows a pattern that drifts left in the rotating frame, showing that the observed saddle shape rotates backwards relative to the rotating frame.  In sequence 2, the flat state was never observed and the sheet transitioned directly from draped rotation (see Fig. \ref{figsetup}(e)$i,ii$) to a whirling state (see Fig. \ref{figsetup}(f)$i,ii$) as $\Omega$ was increased. In Fig. \ref{figsetup}(e)$ii$, the draped rotation pattern is also seen to exhibit a slight retrograde rotation. This behavior was noted for heavier sheets which also showed higher mode shapes during draped rotation (see $m=4$ in Fig. \ref{figsetup}(e)$i$). See SI video for representative experimental videos of the three states.

Before turning to quantify the transitions between the states, we recognize that there are five dimensionless parameters: an aspect ratio $\epsilon=h/R$, an annulus ratio $\alpha=R_i/R$, the ratio of inertial/centrifugal to bending forces $So=\rho h \Omega^2 R^4/D$, the ratio of gravitational to bending forces $Gr = \rho g h R^3/D$, and Poisson's ratio $\nu$. Aerodynamic coupling introduces two additional dimensionless parameters: the fluid loading parameter $\rho_aR/(\rho h)$ and the Reynolds number $Re=\Omega R^2/\nu_a$, where $\rho_a$ and $\nu_a$ are the density and kinematic viscosity of air, respectively. Of these, only three are important in delineating the qualitative transitions that we see: $\epsilon, Gr, So$. We note that the ratio $So/Gr=\Omega^2R/g$ represents the ratio of centrifugal to gravitational acceleration, so that we expect the transition from draping to flattening to occur at $(\Omega_c^{\rm{f}})^2R/g=c_1$, i.e. $So_c^{\rm{f}} = c_1 Gr$, independent of $\epsilon$ (See the Supplementary Information Fig. S2).

In Fig. \ref{figcrit}(a), we plot the ratio of the angular speed of the pattern to the shaft speed $\Omega_s/\Omega$ as a function of $So$, with the lines plotted as a guide for the eyes.  At low $So$, $\Omega_s/\Omega \le 1$; as we increase $So$, $\Omega_s/\Omega \approx 1$ and the pattern on the sheet becomes axisymmetric and the sheet flattens. 
The value of $\Omega_s/\Omega$ starts out smaller for sequence 2 at low $So$ (see most green data in Fig. \ref{figcrit}(a)). As $So$ is increased, $\Omega_s/\Omega$ does not increase to 1 indicating that the sheet never flattens. In both sequences, $\Omega_s/\Omega$ exhibits a sudden decrease marking a significant retrograde rotation of the shape as the sheet transitions to the whirling state (cf. Fig. \ref{figsetup}). The data shown in Fig. \ref{figcrit}(a) are from experiments where $So$ was increased systematically. The transition between these regimes when decreasing $So$ occurs at slightly different values indicating hysteresis. Two representative cases of hysteresis loops are shown in the inset of Fig. \ref{figcrit}(a).

\begin{figure}[t]
\begin{center}
\includegraphics[width=0.45\textwidth]{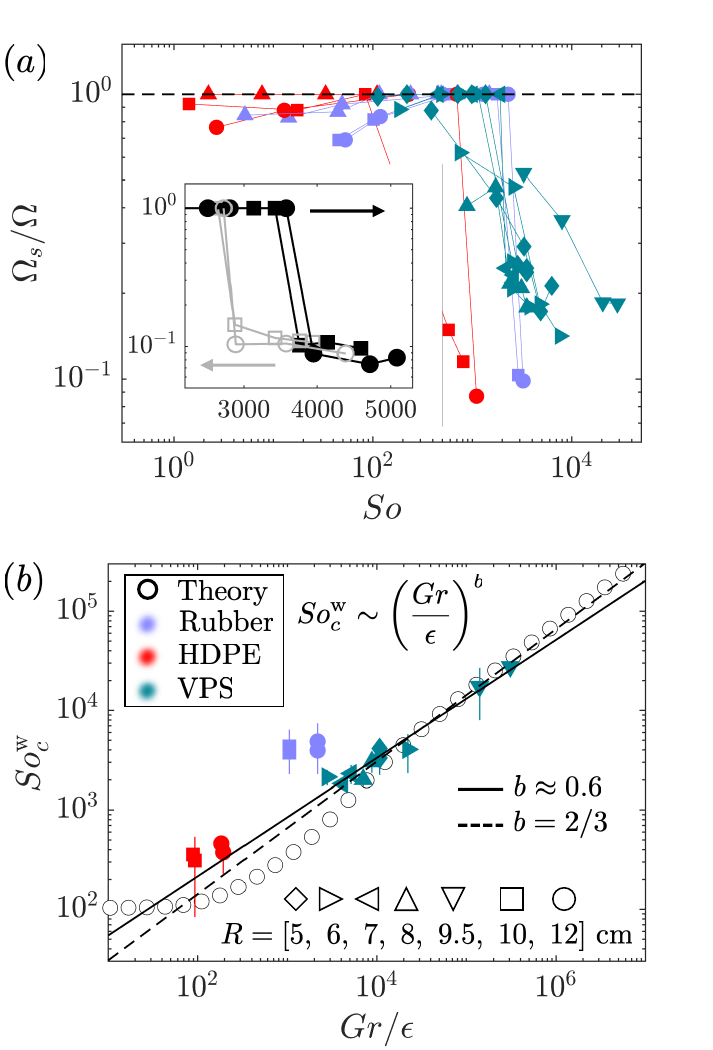}
\caption{\label{figcrit} Characterizing the transition to whirling. (a) The ratio of shape speed to shaft speed $\Omega_s/\Omega$ is plotted as a function of $So$. The inset shows the hysteresis loop where black data corresponds to increasing $So$ and gray data corresponds to decreasing $So$. (b) The critical whirling condition follows a power law $So_c^{\rm{w}} = c_2 (Gr/\epsilon)^b$ for large $Gr/\epsilon$. The color markers are the experimental results where the fitted exponent for data with $Gr/\epsilon>2.5\times10^3$ is $b\approx 0.60$ with $R^2=0.99$ (black line). The white circles are the theoretical prediction from linear perturbation analysis, multiplied by a prefactor of $\approx 5.5$ to align with the experimental results. The black dashed line is the $2/3$ scaling law for reference with $c_2 = 6.6$. The shape of the markers correspond to the $R$ of the sheet and the color of the markers correspond to the material, as shown in the legend. }
\end{center}
\end{figure}

To understand the transition to retrograde whirling, we plot the critical $So$ from our experiments at which whirling occurs, $So_{c}^{\rm{w}}$, in Fig. \ref{figcrit}(b). In the context of rotating flat plates where gravitational effects are neglected ($Gr=0$), the critical speed of whirling is defined as the angular speed at which the backward-traveling wave becomes stationary in the lab frame, at which point the plate becomes susceptible to unbounded resonant growth from any steady lab frame disturbances. This critical speed scales directly with the natural frequency of the lowest frequency bending oscillations of the non-rotating plate \cite{genta2005dynamics} and is given by  $\Omega_{c,flat}^{\rm{w}} \sim (\frac{D}{\rho h R^4})^{1/2}$, which in dimensionless terms corresponds to $So_{c,\rm{flat}}^{\rm{w}}=\textrm{constant}$. In Fig. \ref{figcrit}(b), the colored markers shows the experimental $So_c^{\rm{w}}$  varies almost two orders of magnitude and is clearly not a constant. We now show that this large variation is a consequence of gravitational effects; when $Gr \neq 0$  the frequency of \textit{soft} whirling is one where the critical angular velocity is effectively renormalized by the geometry generated by gravity.

To isolate the role of gravity-induced curvature in determining the intrinsic whirling threshold, we first neglect coupling to the surrounding fluid \footnote{Although the fluid loading parameter $\mathcal\approx0.04$--$0.35$ is not negligible, the unsteady inertial fluid loading vanishes at the stationary-wave resonance in quiescent air. Furthermore, viscous stresses are small at the experimental Reynolds numbers ($Re\sim10^3$--$10^5$). We therefore neglect aerodynamic effects in determining the elastic resonance, but revisit them in estimating the propagation of finite-amplitude whirling waves.} and consider the conservative dynamics of a rotating gravitationally-loaded elastic shallow shell~\cite{love2013treatise}. This approximation allows us to identify the stationary-wave resonance of the elastic continuum, while aerodynamic forces may determine the actual onset, amplitude and propagation speed of the observed whirling state. Linearizing small out of plane perturbations $\hat w'(\hat{r}, \theta, \hat{t})$ about a steady, gravitationally sagged axisymmetric base state $\hat{w}_0(\hat{r})$ as $\hat w = \hat w_0 + \hat w'$, the dynamical equation for $\hat w'$ has the following dimensionless form in the rotating frame (see SI for details), 
\begin{subequations}\label{eq:nondim-governing}
\begin{align}
\frac{\partial^2 \hat w'}{\partial \hat{t}^2}
&
+ \hat{\mathcal{L}}_{\mathrm{s}} \hat w' = 0,
\label{eq:nondim-governing-a}\\
\hat{\mathcal{L}}_{\mathrm{s}} \hat w'
&=
\begin{aligned}[t]
&\hat{\nabla}^4 \hat w'
- So\,\hat{\nabla}\cdot
\left(\mathbf{\hat{N}}\,\hat{\nabla} \hat w'\right)\\
&\quad
+ \hat C_s\,\hat L\big(\hat w_0,\hat\nabla^{-4}\hat L(\hat w_0,\hat w')\big),
\end{aligned}
\label{eq:nondim-governing-b}
\end{align}
\end{subequations}
where $\mathbf{\hat{N}}$ is the in-plane stress tensor of the base state, $\hat L$ is the von K\'arm\'an bilinear operator, and the hats represent dimensionless quantities rescaled as follows: $\hat{w}=w/h,~\hat{r}=r/R,~\hat{t}=t/\sqrt{\rho h R^4/D},~\hat{\mathbf{N}}=\mathbf{N}/\rho h \Omega^2 R^2$. The first term in Eq. (\ref{eq:nondim-governing-b}) is the usual bending term in the linearized limit, the second corresponds to contribution from the in-plane stress projected in the normal direction, and the final term with the coefficient $\hat{C}_{\text{s}}=12(1-\nu^2)$, is the shallow-shell coupling that captures the effect of the gravitational sag and is quadratic in the base state curvature, scaling as $\hat{C}_{\mathrm{s}}\,\hat{\kappa}_0^2 \hat w' \propto [Gr/(\epsilon ~ So)]^2 \hat w'$ (see SI for details). Substituting the normal mode ansatz $\hat w' = \phi_m(\hat{r}) e^{i(m\theta - \hat{\omega}\hat{t})}$ into Eq. (\ref{eq:nondim-governing}) and imposing the condition that the backward-traveling wave becomes stationary in the lab frame gives the whirling condition $m^2 So_{c}^{\rm{w}} = \lambda$ where $\lambda=\langle\phi_m,\mathcal {\hat L}_s\phi_m\rangle/\langle\phi_m,\mathcal \phi_m\rangle$ is the lowest eigenvalue of the $\mathcal {\hat L}_s$ operator for azimuthal wavenumber $m$. We numerically solved for $So_c^{\rm{w}}$ for various $Gr$ and $\epsilon$ ($m=2,~\nu=0.5,~\alpha=0.2$) and the results are shown as white circles in Fig. \ref{figcrit}(b). We see that for  $Gr/\epsilon \ll 10^2$, the onset of retrograde whirling is independent of this parameter, but shows a marked dependence when $Gr/\epsilon \gg 10^2$. In experiments, only $m=2$ (saddle) mode was observed at steady-state. Poisson's ratio was nearly constant across materials ($\nu\approx0.46-0.5$) and although $\alpha$ ranged over $[0.13,~0.32]$, the prefactor only varied $\pm15\%$ of its mean. 

\begin{figure*}[t]
\begin{center}
\includegraphics[width=0.99\textwidth]{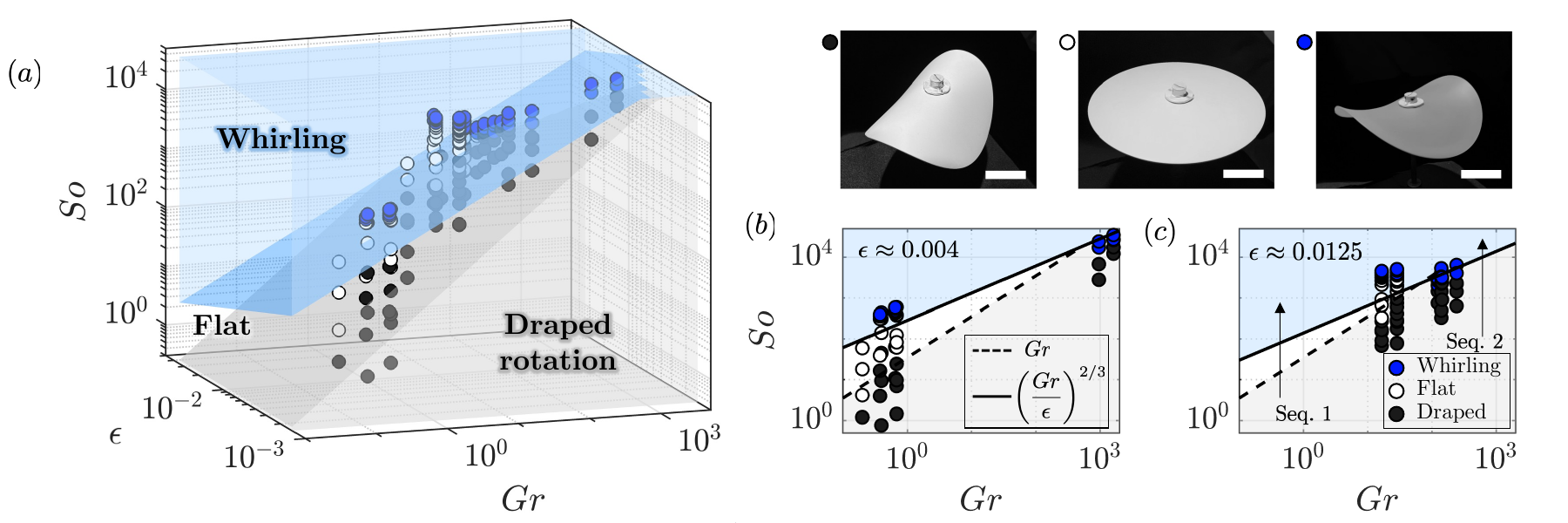}
\caption{\label{figphase} Phase diagram summarizing the rotating states. (a) Phase diagram as a function of $So,~Gr$ and $\epsilon$ summarizing the three states of a rotating soft sheet: draping, flat, and whirling. The transition from draped to flat is described by the surface $So= c_1 Gr$. The transition from flat to whirling is described by the surface $So=c_2 (Gr/\epsilon)^{2/3}$. Representative experimental images of the three states are also shown, with the scale bars representing 5 cm. (b) Cross-sectional view of the phase diagram at $\epsilon \approx 0.004$. (c) Cross-sectional view of the phase diagram at $\epsilon \approx 0.0125$. The constants $c_1=35$ and $c_2=6.6$ for the surfaces shown in the plots.}
\end{center}
\end{figure*}

To provide a physical explanation of this phenomenon, we note that the eigenvalue problem described by Eq. (\ref{eq:nondim-governing}) has the approximate structure $\lambda \approx \mu_B + So~\mu_{c} + (\frac{Gr}{\epsilon So})^2 \mu_s$ where $\mu_B,~\mu_c,~\mu_s$ are the modal projection constants of the bending, in-plane stress, and shell terms, respectively. Substituting this into the whirling condition and rearranging, we get
\begin{equation}
(So_c^{\rm w})^2~\mu_B + \bigg(\frac{Gr}{\epsilon}\bigg )^2 \mu_s \approx (m^2-\mu_c)(So_c^{\rm w})^3.
\label{approxEigen}
\end{equation}
Here, the term on the right hand side represents the net rotational contribution after accounting for centrifugal stiffening, with resonance requiring $m^2>\mu_c$. For low values of gravity, $Gr\rightarrow0$, and when bending dominates, we obtain the scaling $So_{c,\rm{flat}}^{\rm{w}}=c_1=\textrm{constant}$. In contrast, when gravity dominates bending, the sheet is bent and the dominant balance is between the second term on the left side and the term on the right hand side, so that
\begin{equation}
So_{c}^{\rm{w}} = c_2\left( \frac{Gr}{\epsilon} \right)^{2/3},
\label{eq:whirling_scaling}
\end{equation}
corresponding to the balance between the curvature-induced stiffness and the net rotational contribution in Eq. (\ref{eq:nondim-governing-b}).  Introducing the  gravitational frequency scale  $\omega_g = (g/h)^{1/2}$, Eq.~(\ref{eq:whirling_scaling}) yields
\begin{equation}
\Omega_{\mathrm{c}}^{\mathrm{w}}
= \sqrt{c_2}\,\left(\Omega_{\mathrm{c,flat}}^{\mathrm{w}}\right)^{1/3}\omega_g^{2/3}
= \sqrt{c_2}\left(\frac{E g^2}{12(1-\nu^2)\rho R^4}\right)^{1/6}.
\end{equation}
We see that gravity transforms the classical whirling 
frequency into a weighted geometric mean of elastic and 
gravitational frequencies, while making it independent of 
sheet thickness in the gravity-dominated limit \footnote{The crossover between these regimes follows from Eq.~(\ref{approxEigen}). Defining $\Pi=m^2-\mu_c$, $So_0=\mu_B/\Pi$,$X=So_{c}^{\rm{w}}/So_0$, and $\Lambda=(\mu_s/\Pi So_0^3)^{1/2}Gr/\epsilon$, we obtain $X^2(X-1)=\Lambda^2.$ For approximately constant modal coefficients,  $X-1\sim\Lambda^2$ when $\Lambda\ll1$, while $X\sim\Lambda^{2/3}$ when $\Lambda\gg1$. This crossover embodies the transition from intrinsic bending stiffness to gravity-induced geometric stiffness in determining the whirling resonance.}. Fundamentally, we would like to emphasize that this is closer to a resonance---a backward wave stationary in the laboratory frame---rather than a dynamical instability. Interestingly, the wrinkled-to-axisymmetric transition of transversely loaded spinning membranes obeys the same scaling \cite{simmonds1962finite,delapierre2018wrinkling}, because the balance between the in-plane stretching induced by the gravitational sag and the centrifugal stresses are governed by the same dimensionless parameters. Nevertheless these are distinct physical phenomena: unwrinkling is a static instability ($\lambda=0$), after which the sheet simply becomes axisymmetric, whereas whirling arises via a resonant excitation arising from the axisymmetric state ($\lambda=m^2 So_{c}^{\rm{w}}$).

The scaling law in Eq. (\ref{eq:whirling_scaling}) is plotted in Fig.~\ref{figcrit}(b) (dashed black line) and is consistent with the measured exponent of $0.60$ (95\% CI: 0.50--0.70, $R^2=0.99$) for $Gr/\epsilon>2.5\times10^3$. At smaller $Gr/\epsilon$, both theory and experiments approach the classical flat-plate limit of constant $So_c^{\rm{w}}$. While capturing this crossover, the theory underestimates the whirling threshold by a factor of $\approx5.5$ in the gravity-dominated regime, with larger discrepancies for HDPE sheets. This may reflect the neglect of fluting and stretching-induced stresses (see SI).  

Moreover, finite-amplitude whirling requires energy transfer from the driven rotation to the wave, as in aerodynamically coupled vibrations of spinning disks~\cite{mote1993aerodynamically,renshaw1994aerodynamically}. This may explain the delayed onset, particularly for nearly rigid sheets, while the threefold higher $So_c^{\rm w}$ of rubber relative to VPS sheets at comparable $Gr/\epsilon$ suggest additional material effects. In all cases, whirling patterns propagate retrograde in the rotating frame, with $\Omega_s/\Omega\approx0.1$--$0.3$ (Fig. \ref{figcrit}(a)).
A heuristic estimate of the pattern speed follows by balancing the aerodynamic drive and drag. The shear stress in the rotating boundary layer of thickness $\delta\sim(\nu_a/\Omega)^{1/2}$, projected onto a corrugation of slope $mA/R$, yields a driving pressure $p_{\rm drive}\sim\rho_a mA(\nu_a\Omega^3)^{1/2}$, while the normal motion of the pattern generates a pressure drag $p_{\rm drag}\sim\rho_a(mA\Omega_s)^2$. Balancing these yields
$\frac{\Omega_s}{\Omega}\sim\left(\frac{R}{mA}\right)^{1/2}Re^{-1/4}.$
For $m=2$, $A/R\sim0.2$ and $Re\sim10^4$--$10^5$ during whirling, we obtain $\Omega_s/\Omega\sim0.09$--$0.2$, comparable to the measured values. Thus, while elastogravitational effects set the whirling resonance, aerodynamic forces may select the wave speed.

Combining the flattening and whirling criteria yields the phase diagram in Fig.~\ref{figphase}(a). With increasing $So$, draped sheets transition to whirling either through an intermediate flat state or directly, depending on the ordering of the thresholds. Equating these thresholds yields $ \epsilon^2 Gr=(c_2/c_1)^3$, implying that the existence of an intermediate flat state is governed by $\rho gR/E$ and is thus independent of sheet thickness in the gravity-dominated limit.  We do not distinguish fluted and axisymmetric draped states owing to limited imaging of the free edge and aerodynamic perturbations, evident in the slow pattern drift in Fig.~\ref{figsetup}(e). This transition has previously been studied in vacuum~\cite{delapierre2018wrinkling}. Cross-sections at two values of $\epsilon$ (Fig.~\ref{figphase}(b) and (c)) show reasonable agreement between the predicted phase boundaries and experiments. Indeed, for a typical medium-weight fabric (see SI), $Gr/\epsilon\approx1.4\times10^8$, yielding $So_c^{\rm{w}}\approx1.8\times10^6$ or a critical whirling frequency of $1.25$ Hz. Thus, a dervish spinning at $0.36$ Hz remains draped, while a Kathak dancer reaching $1.33$ Hz can whirl (Fig.~\ref{figshapes}). No intermediate flat state is predicted since $So_c^{\rm{f}}>So_c^{\rm{w}}$. The phase diagram thus guides the design of sheets with prescribed dynamical shapes at given spin rates.

Inspired by the aesthetic movements in the tunics of dancers, we studied the dynamics of patterns in spinning elastic sheet to illustrate the interplay between centrifugal stabilization and dynamical symmetry breaking: increasing rotation can flatten an initially fluted sheet, only to destabilize it again through retrograde traveling waves. Gravity-induced curvature renormalizes the classical bending frequency into a geometric mean of elastic and gravitational frequencies, independent of sheet thickness in the gravity-dominated limit. Our scaling laws delineate the resulting phase boundaries and illustrate more generally how an external field, by generating geometry, can alter the spectrum and stability of an elastic continuum. Our analysis is limited to a linearized theory and neglects fluid loading, and any nonlinearities - it cannot thus explain the hysteretic transition shown in Fig. \ref{figcrit}(a). Although the added-mass ratio $\rho_aR/(\rho h)\approx0.04$--$0.35$ (for $m=2$) is appreciable, the associated inertial loading vanishes at resonance when the pattern is stationary relative to the ambient air. Above resonance, aerodynamic forces may select the wave speed, as suggested by our scaling estimate. Accounting for nonlinear fluid--structure interactions is a natural next step toward understanding transients and controlling whirling states in dance and elsewhere.

\textit{Acknowledgments:} We thank Yeonsu Jung for guidance on the imaging setup, Emiliano Maldonado for help with the initial experiments, and Benjamin Hafner for help with the Young's modulus measurements. We also thank Aamy Kuldip for providing the images in Fig. \ref{figshapes}(b)-(d) and Elizabeth Parambath for her input. This work was supported by the Simons Foundation and the Henri Seydoux Fund.

\textit{Data availability statement:} The codes to reproduce the numerical results are available at https://github.com/paulkaneelil/rotatingSheets

\bibliography{Experimental_paper/expbib}

\end{document}



\title{Supplementary Information for\\ ``Dancing dervishes, spinning skirts  and whirling waves"}

 \author{Paul R. Kaneelil}
    \affiliation{John A. Paulson School of Engineering and Applied Sciences, Harvard University, Cambridge, MA, 02138, USA}

 \author{Sifan Yin}
    \affiliation{Max Planck Institute for the Physics of Complex Systems, 01187 Dresden, Germany.}
    
 \author{L. Mahadevan}
    \affiliation{John A. Paulson School of Engineering and Applied Sciences, Harvard University, Cambridge, MA, 02138, USA}
    \affiliation{Department of Organismic and Evolutionary Biology, Harvard University, Cambridge, MA 02138, USA.}
    \affiliation{Department of Physics, Harvard University, Cambridge, MA 02138, USA.}

\date{\today}

\maketitle

\section{Description of Supplemental Movies}
\begin{itemize}
    \item SI video 1: Video showing representative experimental results ($Gr\approx17$ and $\epsilon\approx0.02$) for draped rotation state with $\Omega \approx 3.2~\textrm{Hz}$ and $So\approx72$, flat state with $\Omega \approx 12~\textrm{Hz}$ and $So\approx972$, and whirling state with $\Omega \approx 23~\textrm{Hz}$ and $So\approx3600$.
\end{itemize}

\section{Dimensionless Parameters}

Consider a thin circular elastic sheet of outer radius $R$, inner hub radius $R_i$, thickness $h$, density $\rho$, Young's modulus $E$, and Poisson's ratio $\nu$, spinning at constant angular speed $\Omega$ about its central vertical axis and draping under gravitational acceleration $g$, with flexural bending stiffness per unit length $D = \frac{E h^3}{12(1-\nu^2)}$. The primary non-dimensional control parameters that determine the behavior of the sheet are:
\begin{align}
\text{Spin Parameter: } & So = 12(1-\nu^2) \frac{\rho \Omega^2 R^4}{E h^2} = \frac{\rho h \Omega^2 R^4}{D}\label{eq:S0_def} \\
\text{Gravitational Parameter: } & Gr = 12(1-\nu^2) \frac{\rho g R^3}{E h^2} = \frac{\rho g h R^3}{D} \label{eq:Gr_def} \\
\text{Slenderness Ratio: } & \epsilon = \frac{h}{R} \ll 1 \label{eq:eps_def}
\end{align}
The two other parameters are the annulus ratio $\alpha = R_i/R$ and Poisson's ratio $\nu$. 

The parameters $Gr$ and $\epsilon$ are only a function of the details of the material. Considering a dancer's skirt as a medium-weighted fabric with $\rho h = 200$ g/m$^2$, $h=2$ mm, $R = 1.1$ m, and bending stiffness of $D=10^{-5}$ N$\cdot$m \cite{fabricBendingProperties}. This corresponds to $Gr \approx 2.6 \times 10^5,~\epsilon \approx 1.8 \times 10^{-3}$, and therefore $Gr/\epsilon \approx 1.4 \times 10^8$. Note that the outfits used for certain dance forms including kathak are composites that may include multiple layers of fabric and even a weighted edge. Thus, these measurements are an approximation. 

\section{Governing Equations}

In spatial coordinates $(r,\theta)$, time $t$, and out-of-plane deflection $w(r,\theta,t)$, the
non-linear von K\'arm\'an plate equations of motion in the rotating frame for spinning sheets are \cite{nowinski1964,delapierre2018nonlinear}
\begin{equation}
\rho h\frac{\partial^2 w}{\partial t^2}
+ D\nabla^4 w
= \nabla\cdot(\mathbf{N}~\nabla w) + \rho g h,
\end{equation}
\begin{equation}
\nabla^4\Phi = -\frac{E h}{2}\,L(w,w) + 2(1-\nu)\rho h\Omega^2,
\end{equation}
where $\mathbf{N}=(N_r,N_\theta)$ is the in-plane \emph{stress resultant} tensor (force per
length), $\Phi$ is the Airy stress function, and $L(w,\Phi)$ is the von K\'arm\'an bilinear operator defined in polar coordinates as:
\begin{equation}
    \label{vkoperator}
    L(w,\Phi)=\frac{\partial^2w}{\partial r^2}\left(\frac{1}{r}\frac{\partial\Phi}{\partial r}+\frac{1}{r^2}\frac{\partial^2\Phi}{\partial\theta^2}\right)+\frac{\partial^2\Phi}{\partial r^2}\left(\frac{1}{r}\frac{\partial w}{\partial r}+\frac{1}{r^2}\frac{\partial^2w}{\partial\theta^2}\right)-2\frac{\partial}{\partial r}\left(\frac{1}{r}\frac{\partial\Phi}{\partial\theta}\right)\frac{\partial}{\partial r}\left(\frac{1}{r}\frac{\partial w}{\partial\theta}\right).
\end{equation}

We rescale the variables as follows:
\[
\hat r = \frac{r}{R}\in[\alpha,1],\qquad
\hat t = t\sqrt{\frac{D}{\rho h R^4}},\qquad
\hat w = \frac{w}{h},\qquad
\hat{\mathbf{N}} = \frac{\mathbf{N}}{\rho h\Omega^2 R^2},\qquad
\hat\Phi = \frac{\Phi}{\rho h\Omega^2 R^4}.
\]

With these choices, the dimensionless equations of motion become
\begin{equation}
\frac{\partial^2 \hat w}{\partial \hat t^2}
+ \hat\nabla^4 \hat w
= So\,\hat\nabla\cdot(\hat{\mathbf{N}}~ \hat\nabla \hat w)
+ \frac{\mathrm{Gr}}{\epsilon},
\label{main_deflection}
\end{equation}

\begin{equation}
\label{mainstress}
\hat\nabla^4\hat\Phi
= \frac{-6(1-\nu^2)}{So}\,\hat L(\hat w,\hat w)
+ 2(1-\nu) .
\end{equation}

The dimensionless radial and hoop stress resultants are \cite{nowinski1964}:
\begin{equation}
\hat N_r =
\frac{1}{So}\left(
\frac{1}{\hat r}\frac{\partial\hat\Phi}{\partial\hat r}
+ \frac{1}{\hat r^2}\frac{\partial^2\hat\Phi}{\partial\theta^2}
\right)
- \frac{1}{2}\hat r^2,\qquad
\hat N_\theta =
\frac{1}{So}\frac{\partial^2\hat\Phi}{\partial\hat r^2}
- \frac{1}{2}\hat r^2.
\end{equation}

Using these definitions, the in-plane stress term can also be written as
\begin{equation}
\,\hat\nabla\cdot(\hat N_0\hat\nabla\hat w) = \frac{1}{So}\hat L(\hat w,\hat\Phi)
- \left(\frac{1}{2}\hat r^2\hat\nabla^2\hat w + \hat r\,\frac{\partial\hat w}{\partial\hat r}\right).
\end{equation}

\section{Axisymmetric base-state equilibrium}

Under steady rotation $\Omega$ and uniform gravity $g$, the base state is axisymmetric:
$\hat w_0(\hat r)$ and $\mathbf{\hat N_0}=(\hat N_{0,r}(\hat r), \hat N_{0,\theta}(\hat r))$.
The static equilibrium configuration satisfies the 4th-order ODE for deflection and 2nd-order ODE for radial stress:
\begin{equation}
\label{w_highSo}
\frac{d^4\hat w_0}{d\hat r^4}
+ \frac{2}{\hat r}\frac{d^3\hat w_0}{d\hat r^3}
- \frac{1}{\hat r^2}\frac{d^2\hat w_0}{d\hat r^2}
+ \frac{1}{\hat r^3}\frac{d\hat w_0}{d\hat r}
- So\left[
\hat N_{0,r}\frac{d^2\hat w_0}{d\hat r^2}
+ \left(
\frac{\hat N_{0,r}}{\hat r}
+ \frac{d\hat N_{0,r}}{d\hat r}
\right)\frac{d\hat w_0}{d\hat r}
\right]
= \frac{\mathrm{Gr}}{\epsilon},
\end{equation}
\begin{equation}
\label{Nr_highSo}
\frac{d}{d\hat r}\left(
\frac{1}{\hat r}\frac{d}{d\hat r}(\hat r^2\hat N_{0,r})
\right)
= -\frac{6(1-\nu^2)}{So}\frac{1}{\hat r}
\left(\frac{d\hat w_0}{d\hat r}\right)^2
- (3+\nu)\hat r.
\end{equation}
The hoop stress resultant is given by radial force balance:
\begin{equation}
\label{ntheta}
\hat N_{0,\theta}(\hat r) = \hat N_{0,r} + \hat r\frac{d\hat N_{0,r}}{d\hat r} + \hat r^2.
\end{equation}

The boundary conditions associated with the governing equations are a clamped inner edge (at $\hat r = \alpha$) so that,
\begin{equation}
\label{bc1}
\hat w_0(\alpha) = 0,\qquad
\frac{d\hat w_0}{d\hat r}(\alpha) = 0,\qquad
(1-\nu)\hat N_{0,r}(\alpha) + \alpha\frac{d\hat N_{0,r}}{d\hat r}(\alpha) + \alpha^2 = 0,
\end{equation}
and a stress-free outer edge (at $\hat r = 1$) so that,
\begin{equation}
\label{bc2}
\hat N_{0,r}(1) = 0,\qquad
\frac{d^2\hat w_0}{d\hat r^2}(1) + \nu\frac{d\hat w_0}{d\hat r}(1) = 0,\qquad
\frac{d^3\hat w_0}{d\hat r^3}(1) + \frac{d^2\hat w_0}{d\hat r^2}(1) - \frac{d\hat w_0}{d\hat r}(1) = 0.
\end{equation}


In the centrifugal tension regime where rotation-induced pre-stress dominates over non-linear stretching ($So \gg 1$),  the radial stress equation (\ref{Nr_highSo}) simplifies to
\[
\frac{d}{d\hat r}\left(\frac{1}{\hat r}\frac{d}{d\hat r}(\hat r^2\hat N_{\text{cf},r})\right)
= -(3+\nu)\hat r,
\]
consistent with a leading-order balance in (\ref{Nr_highSo}) when using a perturbation expansion in the small parameter $1/So$. This limit gives an analytical solution valid upto order $\mathcal{O}(\hat w_0^2/So)$ as $\mathbf{\hat N_0}=\mathbf{\hat N}_{\text{cf}}+\mathcal{O}(\hat w_0^2/So)$. Solving for the result with the above boundary conditions (\ref{bc1}) and (\ref{bc2}) yields
\begin{equation}
\label{nrcf}
\hat N_{\text{cf},r}(\hat r) = C_1 + \frac{C_2}{\hat r^2} - \frac{3+\nu}{8}\hat r^2,
\end{equation}
where the integration constants $C_1, C_2$ are fixed by
\begin{equation}
\label{nrcf_coeffs}
C_2 =
\frac{\alpha^2}{8}\,
\frac{(1-\nu)(3+\nu) + (\nu^2-1)\alpha^2}
{(1-\nu)\alpha^2 + (1+\nu)},
\qquad
C_1 = \frac{3+\nu}{8} - C_2.
\end{equation}
The hoop stress $\hat N_{\text{cf},\theta}$ can be obtained from Eq. (\ref{ntheta}). Substituting $\hat N_{\text{cf},r}(\hat r)$ into the static sag equation (\ref{w_highSo}), neglecting flexural bending relative to
tension at high spin speeds, and rearranging yields
\begin{equation}
-\,\hat\nabla\cdot\left(\hat N_{\text{cf},r}\hat\nabla\hat w_0\right)
= \frac{\mathrm{Gr}}{\epsilon So}.
\end{equation}

Because the differential operator on the left-hand side contains no remaining physical control parameters, the static sag configuration must factorize into a scalar amplitude times a parameter-free, universal non-dimensional shape
function $f(\hat r)$:
\begin{equation}
\hat w_0(\hat r)
= \left(\frac{\mathrm{Gr}}{\epsilon So}\right) f(\hat r),
\end{equation}
where $f(\hat r)$ depends solely on geometric aspect ratio $\alpha$ and Poisson's ratio $\nu$. 

\section{Linear Perturbation Dynamics}
\label{sec:lin_stab}

We treat the axisymmetric centrifugal-tensioned state as the base state to superimpose linear perturbations. Perturbing the equations (\ref{main_deflection}) and (\ref{mainstress}) as $\hat w = \hat w_0 + \hat w'$, $\hat\Phi = \hat\Phi_0+\hat \Phi'$ gives at the first order:

\begin{equation}
\label{perturb}
\frac{\partial^2 \hat w'}{\partial \hat t^2} + \hat\nabla^4 \hat w'
=
So\,\hat\nabla\cdot(\mathbf{\hat N}_{0}~\hat\nabla \hat w')
- \hat C_s\,\hat L\big(\hat w_0,\hat\nabla^{-4}\hat L(\hat w_0,\hat w')\big).
\end{equation}

We note that the first term on the right hand side of Eq. (\ref{perturb}) is the pre-stress while the second term arises from the perturbation of the in-plane stresses, where $\hat{C}_{\text{s}}=\frac{Eh^3}{D}=12(1-\nu^2)$ and $\hat\nabla^{-4}$ is the inverse biharmonic operator. Two simplifications can be made here to gain intuition for the behavior of the solution to this problem. First, assuming $So \gg 1$, we approximate the base state stress as $\mathbf{\hat N}_{0} \approx \mathbf{\hat N}_{\text{cf}}$. Then, the equation of motion for $w'$ given by (\ref{perturb}) becomes
\begin{subequations}\label{eq:nondim-governing}
\begin{align}
\frac{\partial^2 \hat w'}{\partial \hat{t}^2}
&
+ \hat{\mathcal{L}}_{\mathrm{s}} \hat w' = 0,
\label{eq:nondim-governing-a}\\
\hat{\mathcal{L}}_{\text{s}} \hat w' &= \underbrace{\hat{\nabla}^4 \hat w'}_{\rm bending} - \underbrace{So~\hat{\nabla} \cdot (\mathbf{\hat N}_{\text{cf}}~ \hat{\nabla} \hat w')}_{\rm in-plane~stress} + \underbrace{\hat C_s\,\hat L\big(\hat w_0,\hat\nabla^{-4}\hat L(\hat w_0,\hat w')\big)}_{\rm shell}.
\label{eq:nondim-governing-b}
\end{align}
\end{subequations}
Second, under the shallow-shell simplification, the last term in Eq. (\ref{eq:nondim-governing-b}) scales as $\hat C_s \,\hat \kappa_0^2\, \hat w'$ \cite{calladine1983theory,vella2011wrinkling,vella2012indentation,lazarus2012geometry}. Since the base state curvature is $\hat{\kappa}_0(\hat{r}) = \left(\frac{Gr}{\epsilon So}\right) \left| f''(\hat{r}) + \frac{1}{\hat{r}} f'(\hat{r}) \right|$, this last term is proportional to $[Gr/(\epsilon~So)]^2 \propto \hat w_0^2$. Balancing the last two terms of Eq. (\ref{eq:nondim-governing-b}), we see that $So \sim \hat w_0^2$ which gives $So \sim (Gr/\epsilon)^{2/3}$.

\begin{figure}[t]
\begin{center}
\includegraphics[width=0.4\textwidth]{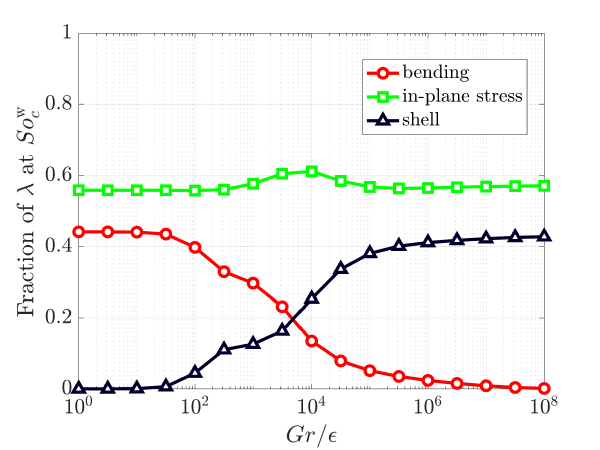}
\caption{\label{figsi_mags} Contributions of the bending (red), in-plane stress (green), and shell (black) terms (see Eq. (\ref{eq:nondim-governing-b})) to the eigenvalue at the whirling threshold shown as fractions of $\lambda$ versus $Gr/\epsilon$, for $m=2$, $\nu=0.5$, and $\alpha=0.2$. Each contribution is the modal projection of the corresponding term in Eq.~(\ref{freqeq}) onto the lowest eigenmode. The in-plane stress fraction stays nearly constant, while the dominant balancing term shifts from bending (small $Gr/\epsilon$) to the shell term (large $Gr/\epsilon$).}
\end{center}
\end{figure}

To go beyond this scaling result, we substitute the modal ansatz $\hat w'(\hat{r}, \theta, \hat{t}) = \phi_m(\hat{r}) e^{i(m\theta - \hat{\omega} \hat{t})}$ into Eq.~\eqref{eq:nondim-governing}. This yields the  equation:
\begin{equation}
\left[ -\hat{\omega}^2  + \hat{\mathcal{L}}_s \right] \phi_m(\hat{r}) = 0.
\label{freqeq}
\end{equation}
Letting $\lambda$ be the lowest spatial eigenvalue of $\hat{\mathcal{L}}_s \phi_m = \lambda \phi_m$ for azimuthal wavenumber $m$, with $\phi_m$ corresponding to the normalized eigenfunction, we get the scalar quadratic frequency equation:
\begin{equation}
\hat{\omega}^2 - \lambda = 0
\end{equation}
Solving for the rotating-frame mode frequencies $\hat{\omega}_{\rm rf}$ exhibits a forward and backward frequency solution:
\begin{equation}
\hat{\omega}_{\rm rf} =  \pm \sqrt{\lambda}
\end{equation}

In the lab frame ($\hat{\omega}_{\text{lf}} = m\sqrt{So} + \hat{\omega}_{\text{rf}}$), the retrograde (backward-traveling) wave frequency is:
\begin{equation}
\hat{\omega}_{\text{lf}}^{(b)} = m\sqrt{So} - \sqrt{\lambda}
\end{equation}
Whirling instability occurs when this backward-traveling wave becomes stationary in the laboratory frame ($\hat{\omega}_{\text{lf}}^{(b)} = 0$):
\begin{equation}
m^2 So_{c}^{\rm w} = \lambda(So_{c}^{\rm w}, Gr, \epsilon)
\label{eq:exact_whirling_cond}
\end{equation}
Because $\hat{\mathcal L}_s$ is self-adjoint and its three terms scale as $1$, $So$ and $(Gr/\epsilon So)^2$, the spatial eigenvalue $\lambda=\langle\phi_m,\mathcal {\hat L}_s\phi_m\rangle/\langle\phi_m,\mathcal \phi_m\rangle$ can be projected onto the lowest mode and has the approximate structure
\begin{equation}
\label{eigenapprox}
\lambda \approx \mu_B + So~\mu_{c} + \bigg(\frac{Gr}{\epsilon So}\bigg)^2 \mu_s, 
\end{equation}
where $\mu_B,~\mu_c, \mu_s$ are the modal projection constants of the bending, pre-stress, and shell terms, respectively. Substituting (\ref{eigenapprox}) into the whirling condition given by (\ref{eq:exact_whirling_cond}) and rearranging, we get
\begin{equation}
(So_c^{\rm w})^2~\mu_B + \bigg(\frac{Gr}{\epsilon}\bigg )^2 \mu_s \approx (m^2-\mu_c)(So_c^{\rm w})^3.
\end{equation}
When gravity is negligible and when bending dominates, this leads to $So_c^{\rm w} \sim \textrm{constant}$, a familiar result for rotating disks and flat plates \cite{lamb1921vibrations,genta2005dynamics}. Conversely, when gravity dominates, we get 
\begin{equation}
So_{c}^{\rm w} \propto \left( \frac{Gr}{\epsilon} \right)^{2/3}.
\label{eq:whirling_scaling_final_supp}
\end{equation}
To test this, we numerically solved Eq. (\ref{eq:exact_whirling_cond}) for the eigenvalues and the results for $So_c^{\rm w}$ are given (white circles) in Fig. 3(b) in the main text. While the prefactor of the numerical result varies by a factor of about 5.5 from the experimental one, the scaling in Eq. (\ref{eq:whirling_scaling_final_supp}) is in agreement over multiple decades in $Gr/\epsilon$. 

In Fig. \ref{figsi_mags} we show the contributions of the bending, in-plane stress and shell terms to the eigenvalue at the whirling threshold, $\lambda = m^2 So_c^{\rm w}$, as a function of $Gr/\epsilon$ for mode $m=2$ ($\nu=0.5$, $\alpha=0.2$). The in-plane stress supplies a nearly constant fraction ($\approx 57\%$), so centrifugal stiffening alone cannot stop the backward wave. The remaining deficit is supplied by bending at small $Gr/\epsilon$, giving $So_c^{\rm w} \approx$ const., and by the shell term at large $Gr/\epsilon$, giving $So_c^{\rm w} \propto (Gr/\epsilon)^{2/3}$, with the crossover near $Gr/\epsilon \sim 10^3$–$10^4$, consistent with the experimental results in Fig. 3(b) in the main text. 

Revisiting our $So \gg 1$ approximation, notice that the magnitude of the neglected term reaches $\mathcal{O}(1)$ at the whirling threshold. Nevertheless, the scaling remains true and the exact solution will only alter the prefactor.

\section{Draped-to-flat transition}
\begin{figure}[h]
\begin{center}
\includegraphics[width=0.5\textwidth]{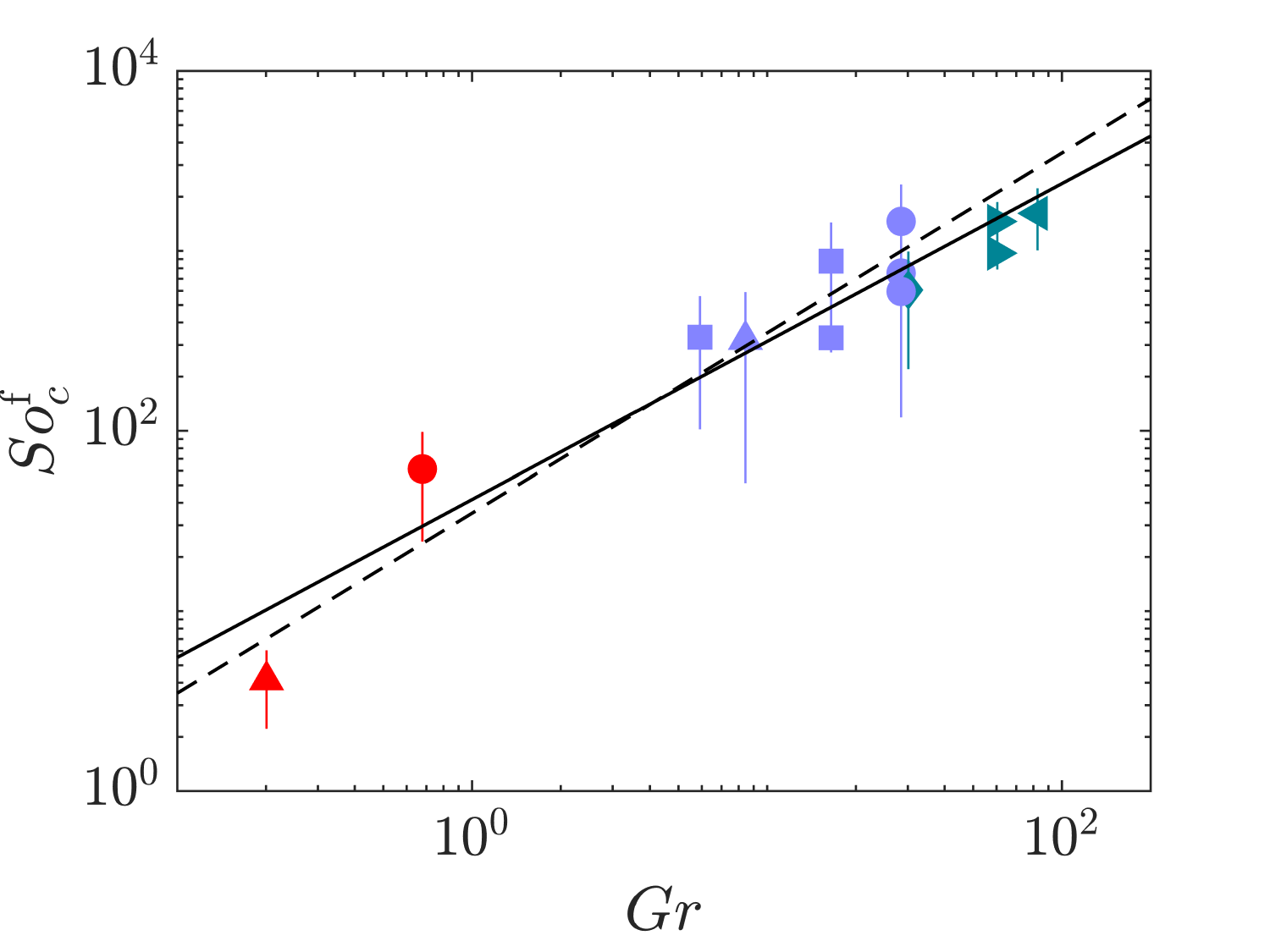}
\caption{\label{figsi_flat} The transition from draping to flat is expected to scale as $So_c^{\rm{f}}=c_1~Gr$. The colored markers represent the experimental data and the black solid line shows the best fit with a scaling law exponent of $0.90$ ($R^2=0.7$). The dashed line shows the expected scaling with $c_1=35$.}
\end{center}
\end{figure}

\textit{Disclosure:} Claude Opus 5.5 (Anthropic) and GPT-6 Luna (OpenAI) were used to assist with symbolic manipulation while deriving the equations, and for writing and debugging the MATLAB code to solve for the critical whirling condition. The authors directed this work and have verified the results independently.

\bibliography{Experimental_paper/expbib}